\documentclass[useAMS,usenatbib]{mn2e}
\usepackage{amsmath,amssymb}
\usepackage{times}
\usepackage{graphicx}
\numberwithin{equation}{section} 
\title[Surface mass density of the Einasto family of dark matter haloes: Are they Sersic-like?]{Surface mass density of the Einasto family of dark matter haloes: Are they Sersic-like?}
\author[Barun Kumar Dhar and Liliya L.R. Williams]{Barun Kumar Dhar$^{1}$\thanks{E-mail:dhar@physics.umn.edu}, Liliya L.R. Williams$^{1}$\thanks{E-mail:llrw@astro.umn.edu}\\
  $^{1}$School of Physics and Astronomy, University of Minnesota, 116 Church Street SE, Minneapolis, MN 55455 USA}
\date{Accepted 29th January, 2010; in original form 30th October 2009}
\begin{document}
  \maketitle
  \label{firstpage}  
  \begin{abstract}
    Recent advances in N-body simulations of dark matter haloes have shown that three-parameter models, in particular the Einasto profile characterized by $d\ln\rho(r)/d\ln r\propto r^\alpha$ with a shape parameter $\alpha \lesssim 0.3$, are able to produce better fits to the 3D spatial density profiles than two-parameter models like the Navarro, Frenk and White (NFW), and Moore et al. profiles.  
  
    In this paper, we present for the first time an analytically motivated form for the 2D surface mass density of the Einasto family of dark matter haloes, in terms of the 3D spatial density parameters for a wide range of the shape parameter $0.1$ $\leq$ $\alpha$ $\leq$ $1$. Our model describes a projected (2D) Einasto profile remarkably well between $0$ and $(3-5)$ $r_{200}$, with errors less than $0.3$ per cent for $\alpha$ $\leq 0.3$ and less than $2$ per cent for $\alpha$ as large as 1. This model (in 2D) can thus be used to fit strong and weak lensing observations of galaxies and clusters whose total spatial(3D) density distributions are believed to be Einasto-like. Further, given the dependence of our model on the 3D parameters, one can reliably estimate structural parameters of the spatial (3D) density from 2D observations.
   
    We also consider a Sersic-like parametrization for the above family of projected Einasto profiles and observe that fits with a Sersic profile are sensitive to whether one fits the projected density in linear scale or logarithmic scale and yield widely varying results. Structural parameters of Einasto-like systems, inferred from fits with a Sersic profile, should be used with caution.
  \end{abstract}     
  \section{Introduction}\label{intro}  
  Gravitational lensing signatures are a response to the projected surface mass column density of matter $\Sigma(\vec R)$ along the line of sight in galaxies and clusters. Upon a suitable deprojection and circular averaging, an estimate of $\Sigma (R)$ can, in principle, be used to trace the spherically averaged 3D density profile $\rho(r)$.
  
  In the past few years, N-body simulations have shown [\cite{P03}, \cite{Nav04} (Nav04), \cite{M06} (M06), \cite{S09}(S09)] that three-parameter models, especially the Einasto \citep{Einasto} profile and the \cite{PS97} de-projected Sersic profiles (PS97), are able to produce better fits to the 3D density profiles of galaxy and cluster-sized dark matter haloes than two-parameter models (\cite{NFW} $\&$ (1996) (NFW), \cite{Moore99}). While the PS97 profile has a well known 2D sky projected form - the Sersic \citep{Sersic} profile, there has been no such analytical counterpart for the Einasto profile. 
  
  In this paper, we present a very good approximation for the 2D projection of the Einasto family of 3D profiles. Thus, if the 3D total mass density is believed to be Einasto-like, our model can be used to parametrically describe the projected 2D surface mass densities of galaxies and clusters in the weak and strong lensing regimes. However, note that even upon radial averaging to smooth out the substructure, not all haloes subscribe to an Einasto profile. For the rest of this paper, we will limit the discussion to the 2D projection of the special case where the 3D profile is Einasto-like.
  
  In 3D, the functional form of the Einasto \citep{Einasto} profile is given by:
  \begin{align}\label{einasto}
    \ln[\frac{\rho(r)}{\rho_s}]= -b[(\frac{r}{r_s})^{\frac{1}{n}} -1] 
  \end{align}
  where, $\rho(r)$ is the 3D (spatial) density at $r$, $n$ (or $\alpha=\frac{1}{n}$) is the shape parameter, $b$ is a function of $n$, $\rho_s$ the spatial density at a scale radius $r_s$ and $\rho(0) = \rho_s e^b$.
  
  In 2D, the Sersic \citep{Sersic} profile, which has been used to describe the projected surface brightness profiles of galaxies, is similar in form to the Einasto profile in 3D \eqref{einasto} and is given by:
  \begin{align}\label{sersic}
    \ln [\frac{\Sigma_S(R)}{\Sigma_{R_E}}] = -q[(\frac{R}{R_E})^{\frac{1}{m}} -1]
  \end{align}
  such that $\Sigma_S(0)$$=$$\Sigma_{R_E}~e^q$ where, $R$ is the projected distance in the plane of the sky, $\Sigma_{R_E}$ is the line of sight projected surface brightness at a projected scale radius $R_E$, which can be defined to be the half-light radius of a Sersic profile under the condition: $q=2m-0.3333+0.009876/m$ (PS97) with $m$ (or $\lambda$$=$$\frac{1}{m}$) characterizing the shape of the Sersic profile.
  
  The parameter $\alpha$$=$$\frac{1}{n}$ in \eqref{einasto} defines the shape of the Einasto profile. In the first ever fits to N-body haloes with a Einasto profile, Nav04 found an average value of $0.172$$\pm$$0.032$ for a wide range of halo masses from dwarfs to clusters. For galaxy-sized haloes \cite{Pr06} found $0.133$$\leq$$\alpha$$\leq$$0.167$. \cite{HW08} observed an evolution of $\alpha$ with mass and redshift ($z$) in the Millennium Simulation ($MS$) of \cite{Spring05} and found $\alpha$$\sim$$0.17$ for galaxy and $\sim$ $0.23$ for cluster-sized haloes. A similar trend is supported by \cite{G08} where $\alpha$$\sim$$0.3$ for the most massive clusters in $MS$. Hence, although in this paper we discuss our approximation to a projected Einasto profile for $0.1$$\leq$$\alpha$$\leq$$1$, particular attention is drawn to the domain $\alpha$$\lesssim$$0.3$; where as we shall show in \S 3, the errors due to our approximation out to $30$$r_{-2}$ ($\sim$$(3-5)$$r_{200}$) are $<$ $0.3\%$.
  
  The parameters $r_s$, $b$ and $n$ in the Einasto profile are not independent. It can be seen, that in terms of a dimensionless length $X=\frac{r}{r_s}$, the logarithmic slope of the density profile is given by:
  \begin{align}
    \beta = \frac{d \ln (\rho/\rho_s)}{d \ln X}= -\frac{b}{n} X^{1/n}
  \end{align}
  Nav04 chooses to define $r_s$ such that  $b/n=2$ (the isothermal value of $\beta=-2$) at $r=r_s$ and hence label $r_s$ as $r_{-2}$ and $\rho_s$ as $\rho_{-2}$. Another approach is to use the convention of M06, requiring $r_s$ to include half the total mass. They quote a numerical estimate of $b$$=$$3n-0.333+0.0079/n$. In this paper we will follow the Nav04 parametrization.
  
  Since the Sersic profile describes the 2D surface brightness profiles of galaxies reasonably well, a natural question is: does the 2D surface mass density of Einasto-like 3D dark matter haloes also follow a Sersic-like description? For the Nav04 simulations, \cite{M05} (M05) have shown that a Sersic function does produce fits with acceptable errors in the range of $r_{conv}$ to $r_{200}$, where $r_{conv}$ is the minimum radius of convergence and $r_{200}$ is the virial radius in the N-body simulations of P03 and Nav04. 
  
  In this paper, we focus on the analytical description of the projected Einasto profile. In order to see how well a projected Einasto profile is described by a Sersic profile, we numerically project \eqref{einasto} and find that a Sersic profile produces acceptable fits only in a limited range of the projected radius $R$. We shall show in \S 3 that this range also depends on whether one fits the numerically projected Einasto profile in linear scale (large errors with increasing R) or log scale (large errors with decreasing R) of density. It is thus evident that the Sersic profile does not give an adequate description of a projected Einasto profile for all $R$.
  
  The choice of Sersic fit (log or linear) may, for example, have possibly strong implications in the strong and weak lensing regimes respectively, yielding incorrect results. It is with this perspective, and the observation that the projection of an Einasto profile (i.e. integral of a Sersic-like function) is not a Sersic-like function but rather a Gamma-like function, we present a derivation of the surface mass density (\S 2.1) for the Einasto profile.
  
  Throughout this paper we will describe 3D spatial parameters as $r$, $r_s$ (or $r_{-2}$), $\rho_s$ (or $\rho_{-2}$), $b$ and $n$ (or $\alpha = 1/n$))and 2D projected parameters as $R$, $R_E$, $q$ and $m$ (or $\lambda = 1/m$). We will also refer to the Sersic profile as $\Sigma_S$, our approximation to a projected 2D Einasto profile as $\Sigma_E$ and a numerically projected Einasto profile as $\Sigma_N$. Further, for this paper, $r_{conv}$ and $r_{200}$ have no physical meaning as such. We note from the Nav04 simulations, that on an average $r_{conv} \sim 5$ per cent of $r_{-2}$, and $r_{200} \sim 8-10$ per cent of $r_{-2}$ for galaxies and $\sim 4$ per cent of $r_{-2}$ for clusters. In this paper we will, for the sake of discussion, refer to $r_{conv} \sim$  $0.05$ $r_{-2}$ and $r_{200}$ as $\sim$ $6$ $r_{-2}$. 
  
  In \S 2 we derive a semi-analytical form \eqref{sigmae} for $\Sigma_E$, which for the Nav04 parametrization of $b=2n$, simplifies to \eqref{sigmae2}. In \S 3 we discuss details of estimating the fit parameters in our expression for $\Sigma_E$, followed by a discussion of errors in our approximation. We then present a comparison between $\Sigma_N$ and the best-fitting Sersic profile $\Sigma_S$. In \S 3.4, we note the remarkable accuracy with which \eqref{sigmae2}, along with \eqref{zeta2param} and \eqref{muparam}, can be used to extract the 3D parameters of Einasto-like systems from its 2D form. 
  
  \section{Surface Mass density of the Einasto profile}
  The Einasto profile has generated considerable interest of late and has been used in recent studies of dark matter haloes (Nav04, Pr06, M06, G08, HW08, S09). For the N-body haloes in Nav04, \cite{M05} have shown that within the limited radial range of $r_{conv} \leq r \leq r_{200}$ and shape parameter $0.12 \leq \alpha \leq 0.22$, a deprojected  Sersic profile also fits the 3D distributions almost as well as the Einasto profile, and a Sersic profile fits the non-parametrically estimated 2D surface densities of dark matter haloes with acceptable errors ($\sim 5$ per cent).
 
  In the following discussion, we argue that if a 3D distribution is Einasto-like, the 2D distribution need not be Sersic-like and provide an analytically motivated functional form for the 2D projection, which can be used to describe the surface mass density of the Einasto family of dark matter haloes subscribing to a wide range of the shape parameter ($1\leq n \leq 10$ or $0.1\leq \alpha \leq 1$) and over a wide radial (projected) range $0 \leq R \leq 6$ $r_{200}$.
  \subsection{General shape of 2D Einasto profile: A short discussion}
  In terms of the line of sight distance ($z$), the surface mass density $\Sigma$(R) can be estimated from: 
  \begin{align}\label{sigmabasic}
    \Sigma(R)= 2 \int^{\infty}_{0} \rho(\sqrt{z^2+R^2}) dz
  \end{align}
  While an exact analytical expression for the integral in \eqref{sigmabasic} for the Einasto profile \eqref{einasto} has so far eluded us, we derive below an excellent semi-analytical approximation.
  
  To intuitively motivate the functional form of $\Sigma_E(R)$ for the Einasto profile \eqref{einasto}, observe that at $R=0$, the integral \eqref{sigmabasic} presents us with an exact solution:
  \begin{align}
    \Sigma_E(0)=\frac{2e^{b}r_s n \rho_s}{b^n} \Gamma[n]
 \end{align}
  \begin{flushleft}
    where, $\Gamma[n]$ is the complete Gamma Function.
  \end{flushleft}  
  Hence, for R$>$0, it is reasonable to expect the integral to depend on terms involving incomplete gamma functions. In fact, for the sake of discussion, one can make a very crude assumption that most of the contribution to the integral in \eqref{sigmabasic} at a given $R$ (especially for $R<r_{-2}$) comes from the region $z>R$. Integrating, from some $\zeta$R to $\infty$ (where $\zeta > 1$), one gets:
  \begin{align}
    \Sigma(R)=\frac{2e^{b}r_s n \rho_s}{b^n} \Gamma[n,b(\frac{\zeta R}{r_s})^{\frac{1}{n}}]
  \end{align}      
  Similarly, the integral for $R>r_{-2}$ will have dominant contributions from terms involving $\gamma[a,x]$; where $\Gamma [a,x]$ and $\gamma [a,x]$ are the upper-incomplete and lower-incomplete gamma functions respectively.
  
  This is quite unlike a Sersic \eqref{sersic} function. Hence, although a Sersic profile may fit a projected Einasto profile in a limited range of $R$ it need not be a very good fit for all $R$.
  \subsection{$\Sigma_E$(R): An analytical approximation of projected Einasto profile}  
  With the 3D spatial distance $r=\sqrt{z^2 + R^2}$, where $R$ is the 2D projected distance in the plane of the sky, and $z$ the line of sight distance from the object to the observer, at any given $R$, one can define 3 regions for the integral in \eqref{sigmabasic} for the Einasto profile \eqref{einasto}: \\
 Region I: $z <R$,  integrating from $z=0$ to $\zeta_1 R$, with $\zeta_1 \leq 1$\\
 Region II: $z > R$, integrating from z=$\zeta_2 R$ to $\infty$, with $\zeta_2 > 1$, and\\
 Region III: $z \sim R$, in a neighborhood $\delta$ between
 $\zeta_1 R$ and $\zeta_2 R$ \\\\
 In Region I: $z<R$, the first term on the right hand side (RHS) of \eqref{einasto} can be written as:
 \begin{align}\label{rhs1_1}
   -b\left(\frac{z^2+R^2}{r_s^2}\right)^{\frac{1}{2n}}= -b\left(\frac{R}{r_s}\right)^{\frac{1}{n}}\left[1+\left(\frac{z}{R}\right)^2\right]^{\frac{1}{2n}}
 \end{align}
 Neglecting 4th and higher order terms in ($z/R$), in the binomial expansion of \eqref{rhs1_1}, the integral of 
 \eqref{sigmabasic} (from $z=0$ to $\zeta_1$R) has an analytical approximation:
 \begin{align}
   \rho_s r_s \exp\left[-b\left(\left[\frac{R}{r_s}\right]^{\frac{1}{n}}
     -1\right)\right]\sqrt{\frac{2n}{b}}\left(\frac{R}{r_s}\right)^{\left(1-\frac{1}{2n}\right)} \notag \\
   \times \gamma\left[\frac{1}{2},\zeta^2_1\frac{b}{2n}\left(\frac{R}{r_s}\right)^{\frac{1}{n}}\right]
  \end{align}  
 In Region II: $z>R$, a binomial expansion of the first term of the RHS of \eqref{einasto} gives us:
 \begin{align}\label{binom2}
   -b(\frac{z}{r_s})^{\frac{1}{n}}(1+(\frac{R}{z})^2)^{\frac{1}{2n}}
    = -b(\frac{z}{r_s})^{\frac{1}{n}} \notag \\
    -b(\frac{R}{r_s})^{\frac{1}{n}}\left[\frac{1}{2n}(\frac{R}{z})^{2-\frac{1}{n}} + \frac{1}{4n}(\frac{1}{2n}-1)
      (\frac{R}{z})^{4-\frac{1}{n}}+...\right] 
 \end{align}
 Fits to N-body simulations with the Einasto profile have so far indicated an $n$$>$$3.0$ ($\alpha$$\lesssim$$0.3$). Hence, observing that the leading contribution comes from the 1st term $b(z/r_s)^{\frac{1}{n}}$, one can drop the remaining terms. This is especially true in our primary domain of interest ($3$$\leq$$n$$\leq 10$ , i.e. $\alpha$$\leq$$0.3$). Even for $1\leq$$n$$\leq 3$, although the approximation is not as good (as it is for $n$$\geq$$3$), it continues to provide better fits than a Sersic profile in the entire range $1$$\leq$$n$$\leq$$10$. With this understanding, the integral from $z=\zeta_2R$ to $\infty$, in Region II can be written as:
 \begin{align}      
   \frac{2e^b r_s n \rho_s}{b^n}\Gamma\left[n,b\left(\frac{\zeta_2R}{r_s}\right)^{\frac{1}{n}}\right]
 \end{align}
 In Region III: $\zeta_1 R \leq z \leq \zeta_2 R$, one can not make the approximations made in regions I and II. However, there exists a point $\epsilon R$ between $\zeta_1 R$ and $\zeta_2 R$, where the mean-value approximation will be valid in a domain $\delta$ about $\epsilon R$. Since the density profile falls rapidly for $R<<r_{-2}$ and gradually for $R>>r_{-2}$, the domain of applicability of the mean-value approximation will be such that $\delta <(\zeta_2-\zeta_1)$ for $R<r_{-2}$ ($ \rightarrow 0 $ as $R \rightarrow 0 $) and tending to $\zeta_2-\zeta_1$ for  $R>>r_{-2}$. Further, it should be obvious that $\zeta_1,\zeta_2$ and $\delta$ will depend on the shape parameter $n$ as well. A function describing $\delta$, with such a property is: 
 \begin{align}
   \delta = (\zeta_2-\zeta_1)[1-\exp(-(R/r_{s})^{\mu})] 
 \end{align}
 \begin{flushleft}
   with $\mu=\mu(n)$
  \end{flushleft}
 The remaining (small) excluded region does not add significantly to the integral (refer to error plots in \S 3).
 
 With this, the contribution to the integral in \eqref{sigmabasic}, from region III- a domain $\delta$ around R - from an application of the mean-value theorem at $\epsilon R$ can be written as:
 \begin{align}\label{region3}
   2 \delta R \rho(z=\epsilon R) = 2\delta R \rho_s
   \exp\left(-b\left[\left(\frac{\sqrt{1+\epsilon^2} R}{r_s}\right)^{\frac{1}{n}}-1\right]\right)
 \end{align}  

 A few important observations are in order. First, neglecting terms in the integrand of region I \eqref{rhs1_1} and II \eqref{binom2}, leads to  over-estimating the integrands in those regions. Second, ignoring the contribution from region III and fitting only for $\zeta_1$ and $\zeta_2$, produces good fits (refer to discussion following \eqref{muparam}) but understandably with a $\zeta_2$ greater than $\zeta_2$ with region III included. $i.e.$ including region III, lowers $\zeta_2$ allowing it to be closer to 1, resulting in a larger contribution from the upper-incomplete gamma function. A simultaneous fit of $\zeta_1, \zeta_2$, $\epsilon$ and $\mu$(in $\delta$) accounts for these excess contributions through a negative sign from region III.
 
 In region I, since $4^{th}$ and higher order terms in $z/R$ are neglected, one can fix $\zeta_1 = 1$. Further, although $\epsilon$ should in principle be estimated, we found it to be a reasonable approximation to fix $\epsilon = \frac{\zeta_2 +\zeta_1}{2}$. This reduces the number of parameters to fit to only two - $\zeta_2$ and $\mu$ - which in turn, during the fitting process, compensates for the approximations made on $\zeta_1$ and $\epsilon$.
  
 With these approximations, the surface mass density $\Sigma_E$(R), for $1 \leq n \leq 10$ with $X=R/r_s$ can be written as:
 \begin{align}\label{sigmae}
    \Sigma_E(R) &=\frac{\Sigma_{E0}}{\Gamma(n+1)}\Bigg[n \; \Gamma \left[n,
	b\left(\zeta_2 X\right)^{\frac{1}{n}}\right] \notag \\
      &  \!+\frac{b^n}{2}\sqrt{\frac{2n}{b}} X^{\left(1-\frac{1}{2n}\right)}\gamma\left[\frac{1}{2}, 
	\zeta^2_1\frac{b}{2n} X^{\frac{1}{n}}\right]
      e^{-b X^{\frac{1}{n}}} \\
      &  \!-\delta b^n X
      e^{-b\left(\sqrt{1+\epsilon^2}X\right)^{\frac{1}{n}}}
      \Bigg]      \notag
 \end{align} 
 \begin{flushleft}    
   where, \[\Sigma_{E0} = \Sigma_E(0)=\frac{2e^{b}r_s\rho_s\Gamma(n+1)}{b^n} \] 
   For the Nav04 parametrization of $\frac{b}{2n}=1$ and by choosing $\zeta_1 =1$, \eqref{sigmae} simplifies to \eqref{sigmae2}; where, after substituting factors like $b/2n$ with $1$ in \eqref{sigmae}, we leave $b$ in the rest of the equation in order to reduce clutter. This gives us:
    \begin{align}\label{sigmae2}
      \Sigma_E(R) =\frac{\Sigma_{E0}}{\Gamma(n+1)}\Bigg[n \; \Gamma \left[n,
	  b\left(\zeta_2 X\right)^{\frac{1}{n}}\right]
	\! \notag \\
	+ \frac{b^n}{2} X^{\left(1-\frac{1}{2n}\right)}\gamma\left[\frac{1}{2}, 
	  X^{\frac{1}{n}}\right]
	e^{-b X^{\frac{1}{n}}}
	\!-\delta b^n X
	e^{-b\left(\sqrt{1+\epsilon^2}X\right)^{\frac{1}{n}}}
	\Bigg]      
    \end{align}     
    with,   
    \[\delta = (\zeta_2-\zeta_1)[1-e^{-(R/r_{s})^{\mu}}]\]    
    \[\epsilon = \frac{\zeta_2 + \zeta_1}{2} \]
    \[\zeta_1 = 1\]
    \[b = 2n\]
    \[X = R/r_{-2}\]
  \end{flushleft}
 and, $\zeta_2$ and $\mu$ are numerically estimated (Fig.1 and Fig.2) for the specific conditions of $b=2n$ and $\zeta_1=1$ to yield:
 \begin{align}\label{zeta2param}
   \zeta_2= 1.1513 +\frac{0.05657}{n} -\frac{0.00903}{n^2}
 \end{align}
 \begin{align}\label{muparam}
   \mu    = \frac{1.5096}{n}+\frac{0.82505}{n^2}-\frac{0.66299}{n^3} 
  \end{align}  
 We would also like to note that based on the accuracy of an approximation needed, one can neglect the contribution from region III (the $\delta$ term) and with $\zeta_1 = 1$ fit only for $\zeta_2$. Although we have not explored a functional relation of $\zeta_2(n)$, we note that the best-fitting $\zeta_2$ varies weakly from $1.2145$ at $n=10$ to $1.2194$ at $n=3.0$ to $1.2424$ at $n=1.0$, with a maximum error within $30\; r_{-2}$ (usually reaching a peak around $r_{200}$) of $0.6\%$ at $n=10$, $1.2\%$ at $n=3$ and $2.2\%$ at $n=1$. Hence, within the current domain of $3<n<8$ from N-body simulations (corresponding to $0.12 < \alpha < 0.3$) one can neglect the $\delta$ term and set $\zeta_1 =1$, and $\zeta_2=1.2176$  at the cost of an error in the range ($0.5\%$ to $1.5\%$).
 
 The parametrizations \eqref{zeta2param} and \eqref{muparam} worked very well even at n=14 and n=0.95. However, we have not tested $n>10$ and $n<1$ values rigorously. If greater accuracy is needed, we recommend fitting for $\zeta_2$ and $\mu$.[Refer to the discussion following \eqref{region3}, on why the $\delta$ term  is negative].  
 \begin{figure}
   \includegraphics[width=84mm]{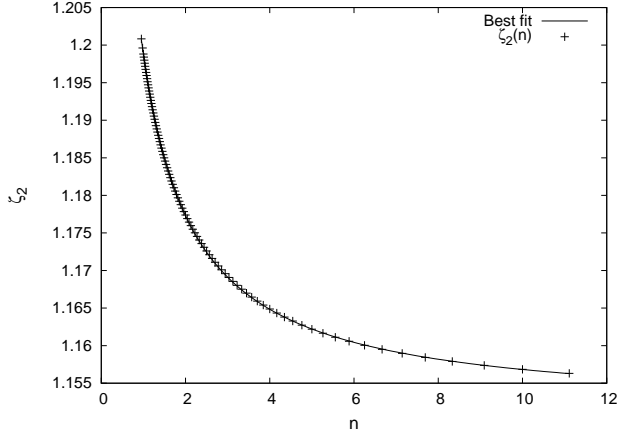}
   \caption{$\zeta_2(n)$ \eqref{zeta2param} for $0.1 \leq \alpha \leq 1.0$ ($1.0 \leq n \leq 10.0$). Each point has been obtained by fitting the numerically projected Einasto profile with \eqref{sigmae2} for $b=2n$ and $\zeta_1=1$.}
    \label{zeta2}
 \end{figure}   
 \begin{figure}
   \includegraphics[width=84mm]{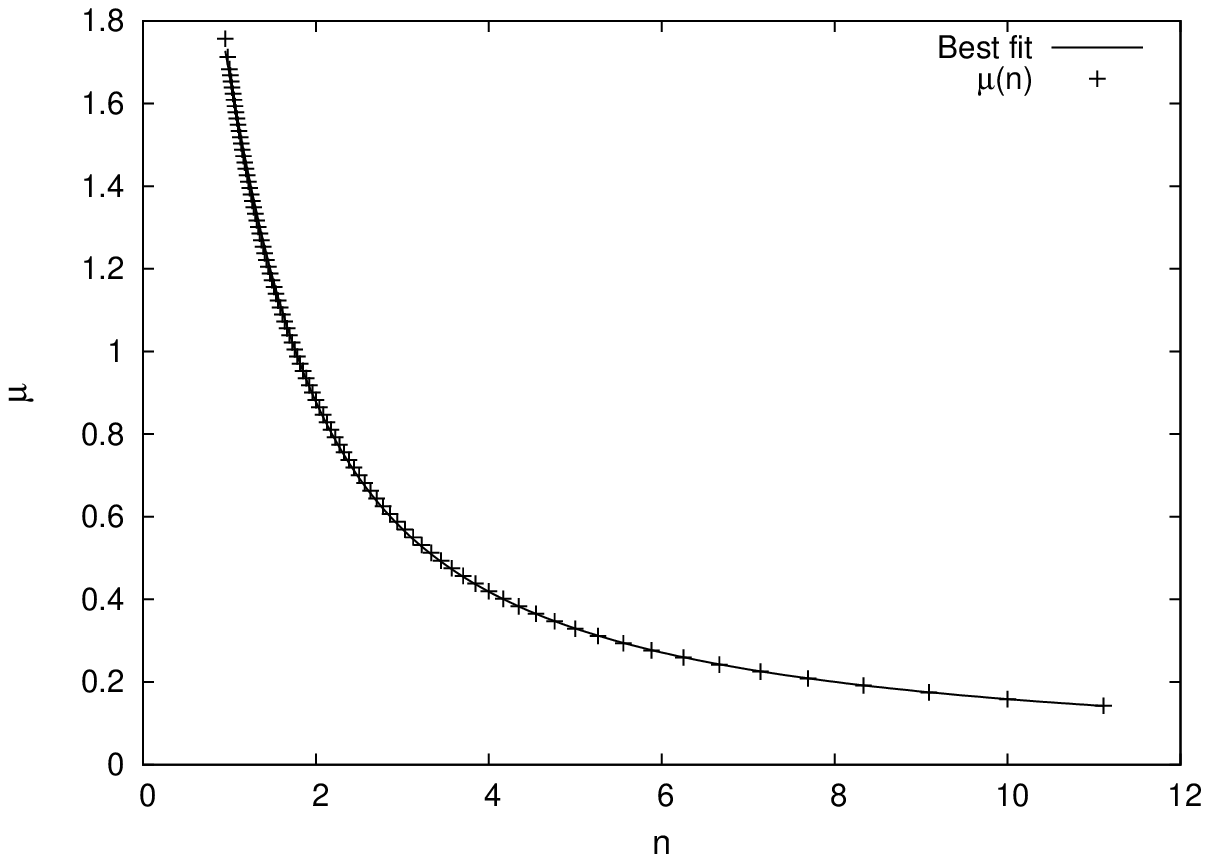}
   \caption{$\mu(n)$ \eqref{muparam} $0.1 \leq \alpha \leq 1.0$ ($1.0 \leq n \leq 10.0$). Each point has been obtained by fitting the numerically projected Einasto profile with \eqref{sigmae2} for $b=2n$ and $\zeta_1=1$.}
   \label{mu}
 \end{figure}       
 This is a useful result because the surface mass density is expressed entirely as a function of the 3D spatial density parameters, which is not the case for any existing projected fitting function. Equation \eqref{sigmae} or \eqref{sigmae2} thus serves two purposes: One, given a 3D Einasto profile, \eqref{sigmae} gives a good approximation to its 2D surface mass density. And two, a good fit to some 2D observations with \eqref{sigmae} for example surface mass density from lensing, will give us the 3D spatial density parameters of a Einasto-like profile. 
 
 This is also quite unlike fitting 2D Einasto profiles with Sersic-like functions, where one first needs to fit for the 2-D and then deproject to fit for the 3D shape parameters (M05), which usually are different without any known existing functional relation between them. 
 \section{Comparison with Numerically Projected Einasto profiles}
 \subsection{Estimating $\zeta_2$ and $\mu$ for $\Sigma_E$}
 We numerically integrate \eqref{sigmabasic} for the profile in \eqref{einasto} and obtain $\Sigma_N(R)$ in the domain $R:(0-30)$ $r_{-2}$ for 90 profiles with a shape parameter in the range $0.1 \leq \alpha \leq 1$. A resolution in $R$ of $0.002$ $r_{-2}$ ($\sim$ $0.05$ $r_{conv}$), allows us to quantify errors due to our approximation in a domain $R <<r_{conv}$, and we report comparison of errors up to $30$ $r_{-2}$ or up to a R where $\Sigma_N(R) \sim 10^{-8} \Sigma (0)$ whichever is earlier.
 
 With $b=2n$ and $\zeta_1=1$, we use a non-linear least squares Levenberg-Marquardt algorithm to estimate the best-fitting values for $\zeta_2$ and $\mu$ by fitting \eqref{sigmae2} to each of the numerically generated $\Sigma_N$ profiles. We find that $\zeta_2$ \eqref{zeta2param} and $\mu$ \eqref{muparam} are best described by a second and third degree polynomial respectively, in $\alpha = 1/n$ (Fig.1 and Fig.2).
 \subsection{A detailed look at errors in the $\Sigma_E$ approximation}  
 Fig. 3 and Fig. 4 describe the fractional error profile between our model $\Sigma_E$ and the numerically projected Einasto profile $\Sigma_N$(R) for a wide range of the shape parameter $\alpha=1/n$. Fig. 3 is more relevant to current N-body simulations, where in $\alpha$ seems to be in the range 0.1 to 0.25. It is worth noting here, that for $\alpha$ as high as 0.25 ($n=4.00$), the largest errors are $<0.3\%$ in the range (0 to 30) $r_{-2}$.
 \begin{figure}
    \includegraphics[width=84mm]{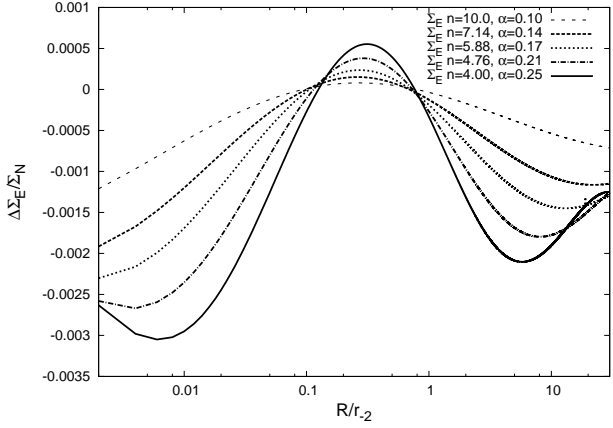}
    \caption{Fractional Error between $\Sigma_E$ and $\Sigma_N$ for $0.1 \leq \alpha \leq
      0.25$, $4 \leq n \leq 10 $. The x-axis plotted in log scale is expressed as a ratio of the 2D projected radius 
      $R$ and the 3D scale radius $r_{-2}$ of the corresponding Einasto profile. $r_{200} \sim 6$ $r_{-2}$ while 
      $r_{conv} \sim 0.05$ $r_{-2}$.}
    \label{sigmae1err}
 \end{figure}  
 \begin{figure}
   \includegraphics[width=84mm]{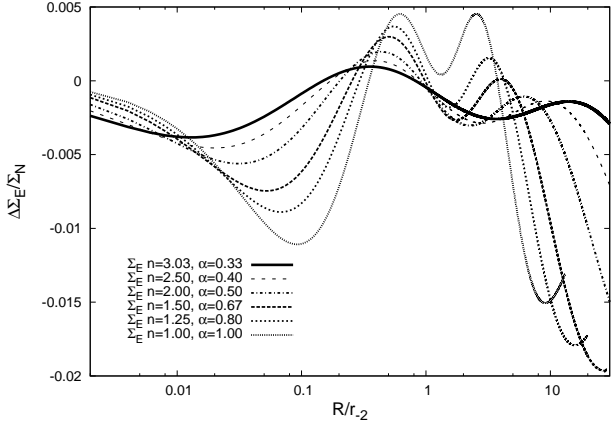}
    \caption{Fractional Error between $\Sigma_E$ and $\Sigma_N$ for $0.33 \leq \alpha \leq
      1.00$, $1 \leq n \leq 3$. The x-axis plotted in log scale is expressed as a ratio of the 2D projected radius 
      $R$ and the 3D scale radius $r_{-2}$ of the corresponding Einasto profile. $r_{200} \sim 6$ $r_{-2}$ while 
      $r_{conv} \sim 0.05$ $r_{-2}$.}     
    \label{sigmae2err}
  \end{figure}  

 In case the range $1 \leq n \leq 3$ becomes relevant in the future, where N-body haloes for $\alpha \gtrsim 0.4$ have not yet been found, we also present in Fig. 4 a comparison between $\Sigma_E$ and $\Sigma_N$. This is also the domain where our assumptions in the $z>R$ region are weaker. Nevertheless, the accuracy of the approximation is striking with the worst error $< 1\%$ within $R < 5$ $r_{-2}$.
 \subsection{Comparison with Sersic profiles}
 In this section we discuss results of fits to $\Sigma_N$ with a Sersic function and superimpose the $\Sigma_E$ model (black solid lines) \eqref{sigmae2} for a relative comparison between $\Sigma_S$ and $\Sigma_E$ (Fig. 5 and Fig. 6). The fits presented here were obtained using log scale of density. One can also obtain fits with density in linear scale; a comparison of residuals for the case $\alpha=0.17$  is presented in Fig. 7. The best fit parameters and consequently the error profile are quite different and is an indication that the best-fitting Sersic profile does not provide an adequate representation of a projected Einasto profile. 
 
 Not evident in the plots (Fig.5-7) are the extremely large errors in the central density $\Sigma_S(0)$ when fitting using log density. For $\alpha=0.1$ (Fig. 5) the relative error at $R=0$ is $375\%$, for $\alpha=0.17$ (Fig.5 and 7), the relative error is $215\%$ and $33\%$ for $\alpha=1$. However, we observed empirically that fits in log density happen to reflect the significance of $R_E$ as enclosing half the total mass of $\Sigma_N$ (it is to be noted that $R_E$ by definition encloses half the total mass of $\Sigma_S$). This is because the domain over which $\Sigma_S$ (with fits obtained in log density) overestimates the density is a relatively small contributor to the total mass and fits with log density in the region $R > r_{-2}$ are good. Since, we can reliably estimate one of the Sersic profile parameters ($R_E$) through fits with log density (as opposed to none in linear scale), we have presented results of fits in log density in Fig. 5 and Fig.6.
 
 The large errors in the central density also have serious consequences for strong lensing. Typically the strong lensing regime extends up to $\sim 0.1$ $r_{-2}$. In strong lensing, image positions correspond to extremum (minima, maxima and saddle) points of the time delay surface. The $j^{th}$ image $\theta_{ij}$ for the $i^{th}$ source $\beta_i$ is given by:
 \begin{align}\label{lenseq}	 
   \vec\theta_{ij} = \vec\beta_i + \frac{1}{\pi}\int \frac{(\vec\theta_{ij} - \vec\theta') \;\kappa(\vec\theta')}{|\vec\theta_{ij} - \vec\theta'|^2}\,d^2 \theta'    
 \end{align}
 where $\kappa(\vec\theta)$ is the normalized surface mass density at an angular position $\vec\theta$. 
 
 Consequently contributions to the integral in \eqref{lenseq} from unusually large density near the center (as a result of fits with a Sersic profile in log density), will produce image separations larger than what one can expect from a numerical projection of the 3D Einasto profile.
 
 Sersic profile fits to model a projected Einasto profile should thus be avoided especially if the central region ($R$$<$$0.1$$r_{-2}$) is being excluded. This is because, as shown in Fig.7, one can get a reasonably good fit (relative error within $10\%$) over a large range of $R$$\gtrsim$$0.1$$r_{-2}$ (fitting in log density) giving an indication that the Sersic profile is a good representation of projected Einasto profile, but doing so will lead to even larger errors in the excluded central region ($R$$<$$0.1$$r_{-2}$). One should thus use caution in interpreting the other structural parameters, the shape parameter $m$ or $\lambda$ and the central density $\Sigma_S(0)$.
 
 Fits in linear density present a different problem. Even though the central errors are much better (relative error $\sim$$10\%$ for $\alpha$$=$$0.17$, Fig.7) than fits using log density, the best-fitting $R_E$ does not enclose half the total mass of $\Sigma_N$ and the errors for large R keep increasing with R.
 
 Although the Sersic profile is not a good representation of a projected Einasto profile, one can fit Sersic profiles in limited domains of R and obtain an estimate of the shape of the projected Einasto profiles in those domains only, but be careful to not use the resulting best-fitting values of $R_E$ and $\Sigma_S(0)$ as a true representation of the half-mass radius and central density of the projected Einasto. In Fig.8, we present two such relations between the 2D Sersic index ($\lambda$$=$$\frac{1}{m}$) and the 3D Einasto index $\alpha$$=$$\frac{1}{n}$ in two domains $R$$<r_{-2}$ and $R$$>r_{-2}$. In these 2 regions, $\lambda$($\alpha$) can be described as power laws. 
 For the domain $R$$<r_{-2}$ we find:
 \begin{align}\label{smallR}
   \lambda(\alpha) = 1.332\; \alpha^{0.741}
 \end{align}
 and for $R > r_{-2}$ defined as in \S 3.1:
 \begin{align}\label{largeR}
   \lambda(\alpha) = 1.037\; \alpha^{0.882}
 \end{align}
 Not only is the shape ($\lambda$) different in the two domains, their evolution with $\alpha$ is also different. Nevertheless, this result can be useful in obtaining an estimate of the shape of projected Einasto profile in these two domains demarcated by $r_{-2}$.
 
The $\Sigma_E$ model, with errors $< 0.5 \%$ does not face any of the above issues. Further, unlike the Sersic profile, the $\Sigma_E$ model can predict the central density with almost $0 \%$ errors due to the existence of an analytical solution. Thus, if the underlying 3D distribution is Einasto-like, the 2D distribution should be modeled with $\Sigma_E$.
  \begin{figure}
    \includegraphics[width=84mm]{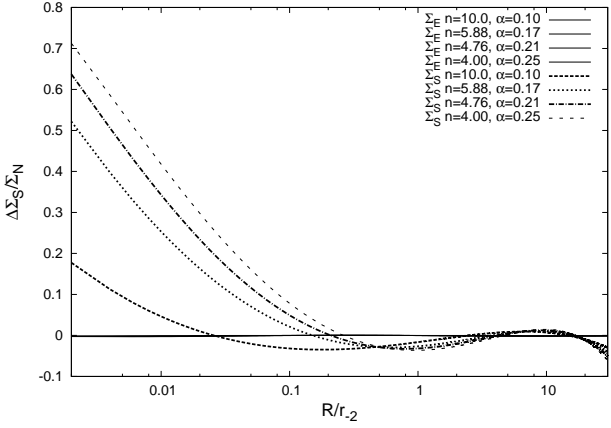}
    \caption{Fractional Error between the best-fitting $\Sigma_S$ and $\Sigma_N$ with Einasto index $0.1 \leq \alpha \leq 0.25$, $4 \leq n \leq 10 $ with fractional error for the best-fitting $\Sigma_E$ (black solid lines $\approx 0$) superimposed. The x-axis plotted in log scale is expressed as a ratio of the 2D projected radius $R$ and the 3D scale radius $r_{-2}$ of the corresponding Einasto profile. $r_{200} \sim 6$ $r_{-2}$ while $r_{conv} \sim 0.05$ $r_{-2}$.}
    \label{Sersicerr1}
  \end{figure}    
  \begin{figure}
    \includegraphics[width=84mm]{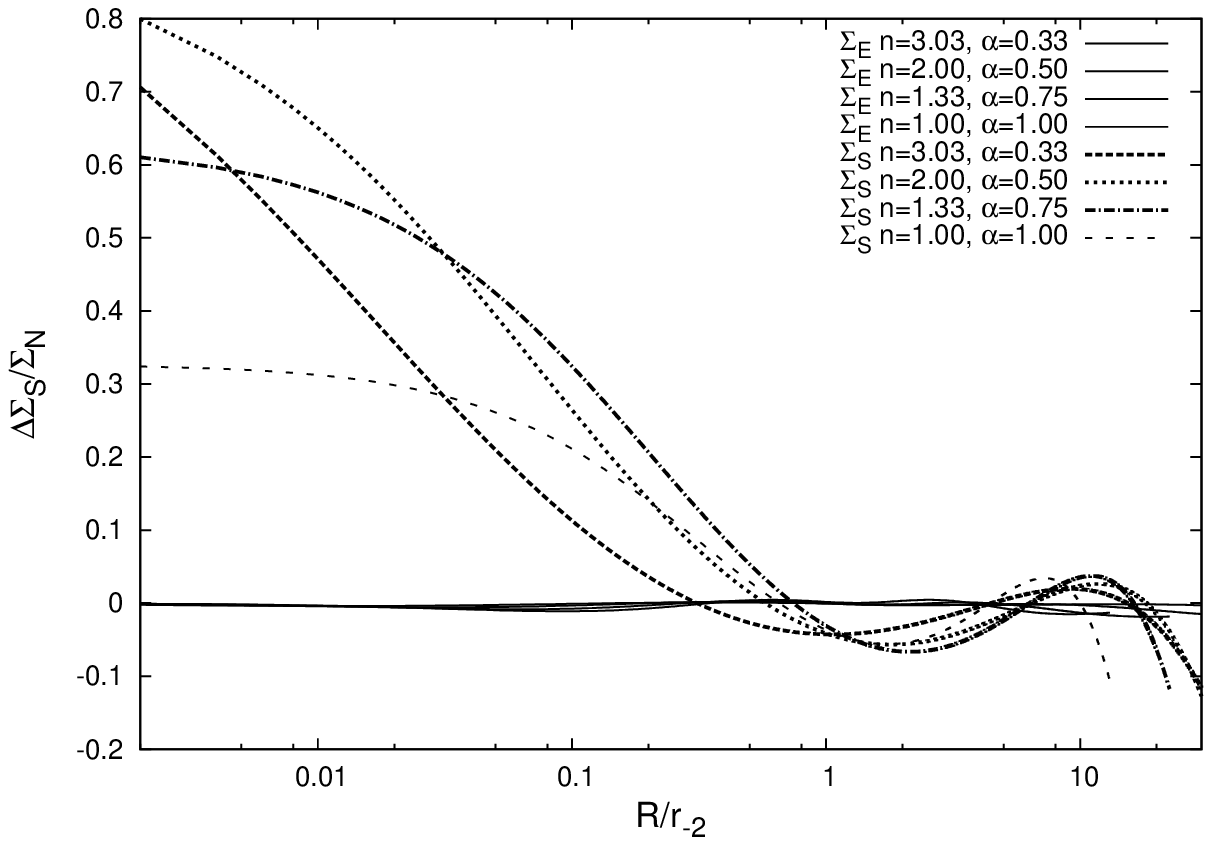}
    \caption{Fractional Error between the best-fitting $\Sigma_S$ and $\Sigma_N$ with Einasto index $0.33 \leq \alpha \leq 1.00$, $1 \leq n \leq 3$ with fractional error for best-fitting $\Sigma_E$ (black solid lines $\approx 0$) superimposed. The x-axis plotted in log scale is expressed as a ratio of the 2D projected radius $R$ and the 3D scale radius $r_{-2}$ of the corresponding Einasto profile. $r_{200} \sim 6$ $r_{-2}$ while $r_{conv} \sim 0.05$ $r_{-2}$.}
    \label{Sersicerr2}
  \end{figure}     
  \begin{figure}
    \includegraphics[width=84mm]{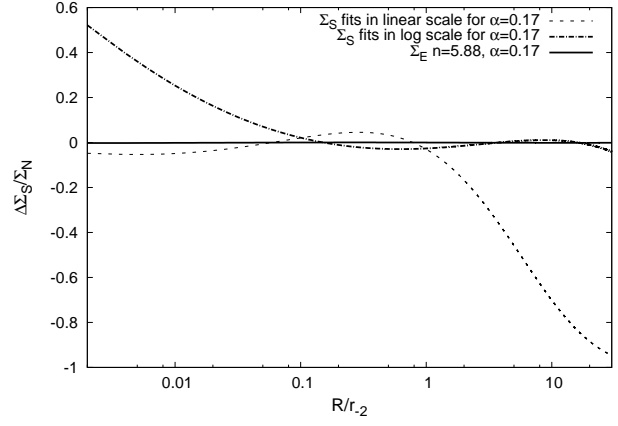}
    \caption{Fractional Error in $\Sigma_S$ from fits to $\Sigma_N$ (for $\alpha=0.17$) with a Sersic profile using log density (dot-dashed line) and a Sersic profile using linear density (dashed line). Not shown in the plots are the errors in central density $\frac{\Delta \Sigma_S(0)}{\Sigma_N(0)}$. The errors are $215\%$ for the fit with log density and $10\%$ for the fit with linear density. The fractional error in $\Sigma_E$ (black solid line $\approx 0$) is superimposed for comparison. The x-axis, plotted in log scale, is a ratio of the 2D projected radius $R$ to the 3D scale radius $r_{-2}$ of the corresponding Einasto profile. $r_{200} \sim 6$ $r_{-2}$ while $r_{conv} \sim 0.05$ $r_{-2}$.}
    \label{Sersiclnlin}
  \end{figure}     
  \begin{figure}
    \includegraphics[width=84mm]{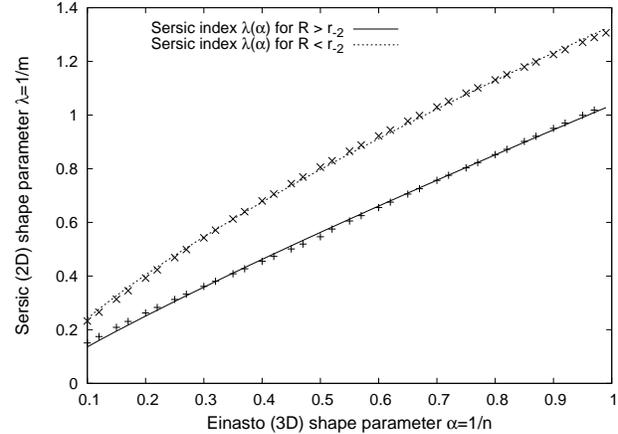}
    \caption{Sersic index($\lambda$=1/m) of a projected (2D) Einasto profile as a function of the 3D Einasto index ($\alpha$=1/n) in two domains $R < r_{-2}$ (dotted) and $R > r_{-2}$ (solid) indicating the absence of a unique Sersic index $\lambda$ for all R. The power law relations for $\lambda$($\alpha$) in equations \eqref{smallR} and \eqref{largeR} are applicable only in these domains.}
    \label{EinSerindex}
  \end{figure}     
  \subsection{Extracting the 3D-parameters from the $\Sigma_E$ model}
  The $\Sigma_E$ model is not just a good description of the projected Einasto profile but is also expressed in terms of the 3D Einasto profile parameters. It should thus be possible to recover the 3D parameters ($\alpha, r_{-2}, \rho_{-2}$) from fits to 2D distributions that subscribe to an underlying 3D Einasto-like system.
  
  For the wide family of numerically projected Einasto profiles $\Sigma_N$ described in this paper, we could recover the 3D parameters for all of them with an accuracy of $\sim 10^{-3}$ or better, by fitting $\Sigma_N$ with \eqref{sigmae2} and the parametrizations of \eqref{zeta2param} and \eqref{muparam} through a non-linear least squares Levenberg-Marquardt algorithm. Given that, as of now robust data (within virialized regions) from N-body simulations are in the domain $r_{conv}$ to $r_{200}$, the fits were performed in this domain. In passing, we note that our results are even better if we fit from $0$ to $30$ $r_{-2}$. Such a high degree of accuracy indicates that if the 2D distribution is indeed a projected Einasto profile, the 3D parameters can be recovered very well even from a limited radial range of observations.
  
  We note that, for the entire range of $0.1 \leq (\alpha=\frac{1}{n}) \leq 1$ the fits always converged for an intial guess of $\alpha < true$ $\alpha$, $r_s > true$ $ r_{-2}$ and $\rho_{-2}$ in the range ($0.1$ to $4$) $\rho^{true}_{-2}$. An inital guess of very low $\alpha \sim 0.05$ and a guess for $r_{-2} \sim r_{200}$ for the type of object (galaxy or cluster) being considered can be a reasonable starting value for the fit to converge. We also did not encounter any local minima. i.e. if the fit converges, it always converged to the true set of ($\alpha$, $r_{-2}$, $\rho_{-2}$).
  \section{Summary and Conclusions}  
  Non-parameteric estimates of density profiles in N-body simulations (Nav04,M06) favour Einasto-like profiles, since they provide better fits than the two-parameter NFW and Moore profiles. \cite{M06} have also shown that a de-projected Sersic profile fits the 3D halo mass distribution almost as well as the Einasto profile, and a Sersic profile provides good fits to non-parametric estimates of surface mass densities (M05) of the Nav04 N-body haloes. 
  
  We have observed that fits with a Sersic function ($\Sigma_S$) to a numerically projected Einasto profile ($\Sigma_N$) are sensitive to whether one fits using linear density $\Sigma_S$ (errors increasing for large R) or log density $ln(\Sigma_S)$ (errors increasing for small R) yielding widely varying results. Consequently, the Sersic profile does not give an adequate description of the projected Einasto profile.

 Sersic profile fits to the surface mass density of N-body haloes (M05), whose 3D spatial densities are well fit by Einasto profiles with $0.12$$\leq$$\alpha$$\leq$$0.22$, have been obtained from the limited radial range of $r_{conv}$ to $r_{200}$. For the haloes in M05 and Nav04, this range is generally less than two decades in radius. Hence, if the 3D distribution is indeed Einasto-like, interpreting structural properties from fits with a Sersic profile, especially $m$ in \eqref{sersic} as the shape parameter, $\Sigma_S(0)$ as the central density and $R_E$ as the half-mass radius, can be misleading.

  In this paper, we have provided an analytical approximation \eqref{sigmae2} to the projected surface mass density of Einasto-like 3D density distributions. The fit errors are well contained to $<$$2\%$ for the projected radial range $0$$\leq$$R$$\leq (10-30)$ $r_{-2}$ equivalent to $(3-5)$$r_{200}$ and shape parameter, $1$$\leq$$n$$\leq$$10$, or $0.1$$\leq$$\alpha$$\leq$$1$. This model can therefore be used both as a fitting function for 2D observations and also to extract the 3D parameters of Einasto-like profiles. Since $\Sigma_E$ fits a projected Einasto profile in a wide radial range, it can be used for fitting strong and weak lensing observations in systems whose total 3D density distribution is believed to be Einasto-like. One can also numerically integrate \eqref{sigmae2} to get reliable estimates of the mass enclosed. 
  
  Finally, we note that the form similarity of \eqref{einasto} and \eqref{sersic}, i.e. fitting functions that describe the 3D mass density of dark matter haloes and the 2D light distributions of galaxies, respectively, could be largely coincidental and should be used with caution when drawing conclusions about the similarity of dynamical evolution that lead to the formation of the stellar components of ellipticals and dark matter haloes.
  \section*{Acknowledgments}
  BKD and LLRW would like to acknowledge the support of NASA Astrophysics Theory Grant NNX07AG86G. We thank Jaan Einasto and Urmas Haud for pointing us to the original literature on the Einasto profile. 

  \label{lastpage}
\end{document}